\documentclass[acmsmall]{acmart}

\renewcommand\footnotetextcopyrightpermission[1]{} % removes footnote with conference information in first column
\usepackage{url}
\usepackage{subcaption}
\AtBeginDocument{%
  \providecommand\BibTeX{{%
    \normalfont B\kern-0.5em{\scshape i\kern-0.25em b}\kern-0.8em\TeX}}}

\copyrightyear{2018}
\acmYear{2018}
\setcopyright{acmlicensed}
\acmConference[Woodstock '18]{Woodstock '18: ACM Symposium on Neural Gaze Detection}{June 03--05, 2018}{Woodstock, NY}
\acmBooktitle{Woodstock '18: ACM Symposium on Neural Gaze Detection, June 03--05, 2018, Woodstock, NY}
\acmPrice{15.00}
\acmDOI{10.1145/1122445.1122456}
\acmISBN{978-1-4503-9999-9/18/06}

\makeatletter
\let\@authorsaddresses\@empty
\makeatother

\begin{document}

%
% The "title" command has an optional parameter, allowing the author to define a "short title" to be used in page headers.
\title{Energy-Aware Scheduling of Task Graphs with Imprecise Computations and End-to-End Deadlines}

%
% The "author" command and its associated commands are used to define the authors and their affiliations.
% Of note is the shared affiliation of the first two authors, and the "authornote" and "authornotemark" commands
% used to denote shared contribution to the research.
%\email{webmaster@marysville-ohio.com}
%\affiliation{%
%  \institution{Institute for Clarity in Documentation}
%  \streetaddress{P.O. Box 1212}

\author{Amirhossein Esmaili}
\affiliation{%
  %\department{Department of Electrical Engineering}
  \department{Department of Electrical Engineering}
  \institution{University of Southern California}
  \city{Los Angeles}
  \state{California}
}
\email{esmailid@usc.edu}

\author{Mahdi Nazemi}
\affiliation{%
  \department{Department of Electrical Engineering}
  \institution{University of Southern California}
  \city{Los Angeles}
  \state{California}
}
\email{mnazemi@usc.edu}

\author{Massoud Pedram}
\affiliation{%
  \department{Department of Electrical Engineering}
  \institution{University of Southern California}
  \city{Los Angeles}
  \state{California}}
\email{pedram@usc.edu}

%\affiliation{%
%  \institution{Inria Paris-Rocquencourt}

%\affiliation{%
% \institution{Rajiv Gandhi University}
 
%\affiliation{%
%  \institution{Tsinghua University}
%  \streetaddress{30 Shuangqing Rd}

%\affiliation{%
%  \institution{Palmer Research Laboratories}
%  \streetaddress{8600 Datapoint Drive}

%\affiliation{\institution{The Th{\o}rv{\"a}ld Group}}
%\email{jsmith@affiliation.org}

%\affiliation{\institution{The Kumquat Consortium}}
%\email{jpkumquat@consortium.net}

%
% By default, the full list of authors will be used in the page headers. Often, this list is too long, and will overlap
% other information printed in the page headers. This command allows the author to define a more concise list
% of authors' names for this purpose.

%
% The abstract is a short summary of the work to be presented in the article.
\begin{abstract}
Imprecise computations provide an avenue for scheduling algorithms developed for energy-constrained computing devices by trading off output quality with the utilization of system resources. 
This work proposes a method for scheduling task graphs with potentially imprecise computations, with the goal of maximizing the quality of service subject to a hard deadline and an energy bound. Furthermore, for evaluating the efficacy of the proposed method, a mixed integer linear program formulation of the problem, which provides the optimal reference scheduling solutions, is also presented. The effect of potentially imprecise inputs of tasks on their output quality is taken into account in the proposed method.
%This work formulates scheduling task graphs with potentially imprecise computations as a mixed integer linear program which seeks to maximize quality of service subject to a hard deadline and an energy bound. 
%
%Furthermore, it introduces a heuristic for efficiently solving the presented optimization problem. 
%
Both the proposed method and MILP formulation target multiprocessor platforms. Experiments are run on 10 randomly generated task graphs. Based on the obtained results, for some cases, a feasible schedule of a task graph can be achieved with the energy consumption less than 50\% of the minimum energy required for scheduling all tasks in that task graph completely precisely.  
%demonstrate energy savings of up to \textcolor{red}{?\%} compared to a solution w ith precise computations in all nodes of task graph. 
%
%Additionally, the proposed heuristic is able to find solutions that are within \textcolor{red}{?\%} of solutions found by mixed integer linear program on the said sample task graphs. \textcolor{blue}{input error}
%

\end{abstract}



%\ccsdesc[500]{Computer systems organization~Embedded systems}

%
% Keywords. The author(s) should pick words that accurately describe the work being
% presented. Separate the keywords with commas.
\keywords{Task Scheduling, Imprecise Computations, Real-time MPSoCs, Input Error}

%
% A "teaser" image appears between the author and affiliation information and the body 
% of the document, and typically spans the page. 
%%  \includegraphics[width=\textwidth]{sampleteaser}
%%  \caption{Seattle Mariners at Spring Training, 2010.}
%%  \Description{Enjoying the baseball game from the third-base seats. Ichiro Suzuki preparing to bat.}

%
% This command processes the author and affiliation and title information and builds
% the first part of the formatted document.
\maketitle
\thispagestyle{empty}

\section{Introduction}
In many real-time applications, it is often preferred for a task to produce an approximate (aka imprecise) result by its deadline rather than producing an exact (aka precise) result late \cite{ASPDAC2008}. 
Imprecise computations increase the flexibility of scheduling algorithms developed for real-time systems by allowing them to trade off output quality with utilization of system resources, such as processor cycles. 

In imprecise computations, a real-time task is allowed to return intermediate and imprecise results of poorer quality as long as it processes a predefined chunk of work that defines its baseline quality. 
The number of processor cycles required for the task to provide this baseline quality is referred to as the mandatory workload of the task. 
Assigning a larger number of processor cycles to a task beyond its mandatory workload leads to an increase in its quality of results. 
In other words, output quality of each task is a monotonic non-decreasing function of processor cycles assigned to it \cite{Greece2010}. 

The workload of a task beyond its mandatory workload is referred to as the optional workload, which can be executed partially. 
The quality of service (QoS) is usually evaluated as a linear or concave function of the number of processor cycles assigned to optional workload of tasks \cite{Greece2010}. When the full workload of a task, both mandatory and optional, is entirely completed, the produced results by that task are considered precise. 

Furthermore, an energy consumption budget may also be one of the main design constraints for energy-constrained computing devices such as embedded systems. 
Some of the prior work have focused on scheduling task graphs with imprecise computations and energy and deadline constraints on single processor platforms \cite{vlsi2006,ASPDAC2008} and multiprocessor platforms \cite{DATE2017,CSE2013}. 
%
%One of the main assumptions in the said prior work is the fact that providing imprecise inputs to tasks within a task graph \textcolor{blue}{do not} affect the quality of their outputs,
In the said prior work, the effect of potentially imprecise inputs of tasks on the their output quality is not considered, and thus QoS can be obtained solely based on the number of processor cycles assigned to optional workload. 
However, in many real-time applications such as video compression or speech recognition \cite{1997} where tasks are interdependent, there is a set of dependent tasks represented by a task graph. Therefore, the input to a task can be dependent on the output of one or more other tasks which may have imprecise results. In the literature, the effect of imprecise input of a task is usually  modeled with an extension in the workload of that task where this extension is responsible for compensating the quality degradation due to imprecise inputs.

This work proposes a method for scheduling  task graphs with potentially imprecise computations,  with the goal of maximizing QoS subject to a hard deadline and an energy bound. The proposed method takes into account the effect of potential extension in the workload of each task based on the quality of its inputs. The proposed method considers task dependencies within a given task graph in order to find tasks (aka nodes) which can be performed imprecisely without having a negative impact on QoS. In addition, for evaluating the efficacy of the proposed method, a mixed integer linear program formulation of the problem, which provides the optimal reference scheduling solutions, is also presented. 
Both the proposed method and MILP formulation target multiprocessor system-on-chip (MPSoC) platforms due to their increasing popularity for many real-time applications. 
To the best of our knowledge, this is the first work that takes into account the effect of workload extension based on input quality in scheduling task graphs on MPSoC platforms where the goal is to maximize QoS subject to a hard deadline and an energy bound. 

The rest of the paper is organized as follows. Section \ref{section.models} explains models used in the paper, formally characterizes tasks with potentially imprecise computations, and presents the problem statement. Next, Section \ref{section.method} explains the proposed method for scheduling task graphs with imprecise computation on an MPSoC platform. It also presents a comprehensive MILP formulation of the same problem, which allows comparing the proposed method with an exact solution. After that, Section \ref{section.results} details experimental results. Finally, Section \ref{section.conclusion} concludes the paper. 

\section{Models and Problem Definition} \label{section.models}
\subsection{Task Model and Imprecise Computation} \label{subsection.task_model}
Tasks to be scheduled are modeled as a directed acyclic graph (DAG) represented by $G(V, E, T_d)$ in which $V$ denotes the set of $n$ tasks, $E$ denotes data dependencies among tasks, and $T_d$ denotes the period of the task graph. $T_d$ acts as a hard deadline for scheduling and each repetition of the task graph should be scheduled before the arrival of the next one.

Each task with the possibility of imprecise computation consists of two parts: a \textit{mandatory part} and an \textit{optional part}. In order for a task to produce an acceptable result, its mandatory part must be completed. The optional part refines the result produced by the mandatory part. If the optional part of a task is not executed entirely, the result of the task is imprecise and the task has an output error. In a task graph, if one or more parent tasks of each task $u$ have an output error, task $u$ will have an input error.
Similar to prior work \cite{Greece2010,end-to-end-QoS}, we assume only the execution of mandatory part of task $u$ will be extended to compensate for the input error and optional part of task $u$ remains the same. This is a valid assumption for many applications such as weather forecasting systems \cite{Greece2010}, image and video processing, and Newton's root finding method \cite{1997}. 
In other words, mandatory part of a certain task can be thought of as the minimum amount of processor cycles required for the task to produce a result with an acceptable quality, and the mandatory part grows when the quality of a task's inputs decrease \cite{end-to-end-QoS}. 
In order for a task graph to be considered feasibly scheduled, at least the potentially extended mandatory workload of each task must be completed before the deadline $T_d$.

The number of processor cycles required to finish the mandatory part of task $u$ when its inputs are error-free is represented by $M_u$. For a task $u$ with nonzero input error, its mandatory workload is extended such that it is capable of producing correct results.  
The number of processor cycles required to process the extension added to $M_u$, which depends on the quality of its inputs, is represented by $M^x_u$. Therefore, the total mandatory workload, represented by $M'_u$, is obtained as follows:
\begin{equation}
M'_u = M_u + M^x_u. \label{equation.TotMan}
\end{equation}

The total optional workload of task $u$, which can be executed partially, is represented by $O_u$. The number of processor cycles actually assigned to the optional workload of task $u$ is represented by $o_u$ ($o_u \leq O_u$).  According to \cite{1997}, the general mandatory extension function of a task can be estimated by a straight line, which provides an upper bound on the amount of required extension. Therefore, the slope of this line, which is represented by $m_u$ and referred to as the task-specific scaling factor \cite{1997,Greece2010}, quantifies the dependency between $E^i_u$ and $M^{x}_u$ as follows:
\begin{equation}
    M^x_u = m_u \times E^i_u, \label{equation.Xscaling}
\end{equation}
in which $E^i_u$ indicates the input error of task $u$. Similar to \cite{Greece2010}, $E^i_u$ in a task graph is defined as follows:
\begin{equation} \label{equation.input_error}
E^i_u=\min\{1, \sum_{j \in par(u)}{E^o_j}\},
\end{equation}
where $par(u)$ is the set of immediate parents of task $u$ and $E^o_j$ represents output error of parent task $j$. $E^o_j$ is defined as the portion of discarded optional workload of task $j$ \cite{1997}, and thus obtained as follows:
\begin{equation} \label{equation.OE}
E^o_j = \frac{O_j-o_j}{O_j}=1-\frac{o_j}{O_j},~ 0 \leq E^o_j \leq 1.
\end{equation}
\iffalse
output error of task $u$, represented by $E^o_u$, is defined as the portion of discarded optional workload of task $u$ \cite{1997}. 
%
Consequently, we have:
\begin{equation} \label{equation.OE}
E^o_u = \frac{O_u-o_u}{O_u}=1-\frac{o_u}{O_u},~ 0 \leq E^o_u \leq 1.
\end{equation}

\textcolor{blue}{The input error of task $u$, represented by $E^i_u$, is modeled as follows: \textcolor{blue}{\underline{discussion should be here}}}
\begin{equation} \label{equation.input_error}
E^i_u=\min\{1, \sum_{j \in par(u)}{E^o_j}\},
\end{equation}
in which $par(u)$ is the set of immediate parents of task $u$.
Typically, a task-specific scaling factor is introduced to quantify the dependency between $E^i_u$ and $M^x_u$ \cite{1997,Greece2010}:
\begin{equation}
    M^x_u = m_u \times E^i_u, \label{equation.Xscaling}
\end{equation}
where $m_u$ is the task-specific scaling factor for task $u$. 
\fi
Based on \eqref{equation.input_error} and \eqref{equation.OE}, we have $0 \leq E^i_u \leq 1$. According to \eqref{equation.Xscaling}, when the input of task $u$ is error-free (i.e. $E^i_u = 0$), $M^x_u = 0$ and thus $M'_u=M_u$. 
On the other hand, when task $u$ has the maximum input-error (i.e. $E^i_u = 1$), $M^x_u = m_u$ and thus $M'_u=M_u+m_u$.
In this case, the mandatory workload extension for task $u$ reaches its maximum. 

Note the assumption that workload extension can always compensate input error is not true in general. However, based on \cite{1997}, we can transform the given mandatory and optional portions of a task workload such that in the worst case, where all transformed optional workloads of parent tasks are discarded, the extension amount obtained by \eqref{equation.Xscaling} would be able to compensate the input error. Therefore, $M_u$ and $O_u$ used in our proposed method are transformed versions of given mandatory and optional workloads of tasks.

The total number of processor cycles assigned to task $u$ is represented by $W_u$ and is obtained as follows:
\begin{equation}
W_u = M'_u + o_u. \label{equation.task_workload}   
\end{equation}

%\textcolor{red}{Note that for a task $u$ with nonzero input error, its mandatory workload is extended such that it is capable of producing correct results. This is achieved by estimating a mandatory workload extension function with two straight lines followed by a linear transformation of the mandatory workload based on the said estimation \cite{feng1997algorithms}. The straight lines provide an upper bound on the extension function.  }
%\textcolor{blue}{In this case, the mandatory workload of task $u$ reaches its maximum.} 
%
      
\subsection{Energy Model} \label{subsection.energy_model}
To model power consumption of a processor when operating at clock frequency $f$, we use the following equation borrowed from \cite{gerards-global}:
\begin{equation} \label{equation.power}
    \rho = \alpha f^{\beta}+{\gamma}f+\delta,
\end{equation}
in which $\rho$ represents the total power consumption, and $\alpha$, $\beta$, $\gamma$, and $\delta$ are the power model constants. $\alpha f^{\beta}$ represents the dynamic power consumption, and $\gamma f+\delta$ represents the static power consumption. 
$\alpha$ is a constant that depends on the average switched capacitance and the average activity factor, and $\beta$ indicates the technology-dependent dynamic power exponent, which is usually $\approx$~3.
Therefore, energy consumption in one clock cycle ($\epsilon_{cycle}$), when executing a task at clock frequency $f$, is obtained \textcolor{black}{from} the following \textcolor{black}{equation}:
\begin{equation} \label{equation.Ecycle}
    \epsilon_{cycle} = \alpha f^{\beta-1}+\gamma+\frac{\delta}{f}.
\end{equation}

\subsection{Problem Statement} \label{subsection.problem_statement}
We seek to schedule a task graph with the possibility of imprecise computations represented by $G(V, E, T_d)$ on a platform comprising of $K$ homogeneous processors in order to maximize QoS subject to a hard deadline and an energy bound. 
Each processor supports a set of $m$ distinct clock frequencies: $\{f_1,f_2,...,f_m\}$. QoS highly correlates with how many processor cycles are assigned to the execution of optional workloads of exit tasks, which are the tasks in the task graph with no child tasks. 
The reason is that the discarded optional workload of tasks other than exit tasks are compensated with extensions in the mandatory workload of their child tasks. 
Consequently, QoS is quantitatively defined as follows: %\textcolor{red}{define a set for exit tasks, write for u in exit tasks under sigma, replace denominator with the cardinality of the set.}
%\begin{equation}
%    QoS = \frac{\sum{P_u\textrm{,}}~\textrm{for u $\in$ exit tasks}}{\textrm{number of exit tasks}},~ 0 \leq QoS \leq 1, \label{equation.QoS}
%\end{equation}
\begin{equation}
    QoS = \frac{\sum_{u \in exit(G)}{P_u}}{|exit(G)|},~ 0 \leq QoS \leq 1, \label{equation.QoS}
\end{equation}
where $exit(G)$ represents the set of exit tasks of task graph $G$, and $P_u$ represents the precision of task $u$. 
$P_u$ is a non-decreasing function of number of processor cycles assigned to the optional workload of task $u$. Similar to \cite{Greece2010}, $P_u$ is defined as follows:
\begin{equation} \label{equation_label}
  P_u=P^T_u+(1-P^T_u)(\frac{o_u}{O_u}),  
\end{equation}
in which $P^T_u$ indicates the minimum precision acceptable from task $u$, aka precision threshold of task $u$. \textcolor{black}{$P^T_u$ assumes values between 0 and 1}. $P^T_u$ indicates the precision of task $u$ when only its (extended) mandatory part is completed \cite{Greece2010}. Based on \eqref{equation_label}, executing only the extended mandatory workload of task $u$ ($o_u=0$) results in $P_u=P^T_u$. On the other hand, executing the entire optional workload of task $u$ ($o_u=O_u$) in addition to its extended mandatory workload leads to $P_u=1$. For other values of $o_u$, $P^T_u<P_u<1$.

\section{Proposed Framework} \label{section.method}
The proposed framework comprises of two main steps:
\begin{enumerate}
    \item determining the number of processor cycles assigned to optional workload of non-exit tasks, and
    \item scheduling tasks on an MPSoC for maximizing QoS subject to energy and deadline constraints.
\end{enumerate}

\subsection{Determining the Number of Processor Cycles Assigned to Optional Workload of Non-Exit Tasks} \label{section.imp_label}
The first step of the proposed method tries to minimize the summation of total workload of non-exit tasks plus the total (extended) mandatory workloads of exit tasks. 
The intuition behind choosing such objective function is the fact that minimizing the total number of processor cycles associated with the aforementioned portions of tasks leads to having more processor cycles available for executing optional workloads of exit tasks as there are fixed deadline and energy budget constraints. This can result in increased QoS according to \eqref{equation.QoS}. 
Therefore, we aim to minimize the following expression:
\begin{equation} \label{equation.bad_workload}
    \left[~\sum_{\textrm{for u $\in$ non-exit tasks}}{W_u}~\right]~~+\left[~\sum_{\textrm{for v $\in$ exit tasks}}{M'_v}~\right].
\end{equation}

We first explain our approach for minimizing \eqref{equation.bad_workload} for two simple task graphs that constitute base cases. Then, we explain our proposed algorithm for a general task graph.

\begin{figure}[b]
\centering
%\captionsetup{justification=centering}
\includegraphics[width=0.45\textwidth]{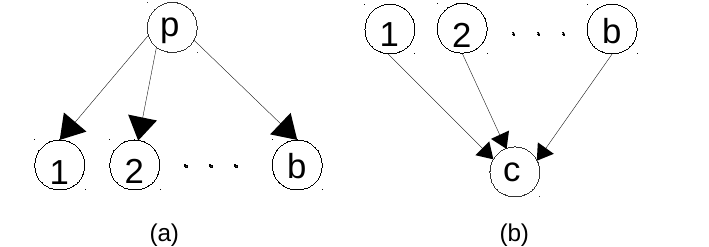}
\caption{\textcolor{black}{ \footnotesize Task graphs of (a) base case 1 and (b) base case 2}} \label{fig.base_cases}
\end{figure}
\textbf{Base Case 1: }Consider the task graph demonstrated in Fig.~\ref{fig.base_cases}a. It consists of a parent task $p$, alongside $b$ child tasks. The workload defined in \eqref{equation.bad_workload} for this simple task graph can be written as follows:
\begin{equation} \label{equation.source_initial}
    [M'_{p}+o_{p}]+[\sum_{i=1}^{b}M_i+\sum_{i=1}^{b}m_i \times (1-\frac{o_{p}}{O_{p}})],
\end{equation}
in which subscripts $p$ and $i$ are used for referring to workload components of the parent task and child tasks in Fig.~\ref{fig.base_cases}a, respectively.
\eqref{equation.source_initial} can be rewritten as:
\begin{equation} \label{equation.source}
    [M'_{p}+\sum_{i=1}^{b}(M_i+m_i)]+[o_{p}\times(1-\frac{\sum_{i=1}^{b}m_i}{O_{p}})].
\end{equation} 

In \eqref{equation.source}, the first term in the summation does not depend on how many processor cycles are assigned to $o_{p}$ (note that the actual workload of $M'_{p}$ depends on the input error of the parent task and not $o_{p}$). 
However, the second term is a function of $o_{p}$ and minimizing this term leads to minimization of \eqref{equation.source}. 
Two possible scenarios are postulated in this case:
\begin{enumerate}
    \item if $\sum_{i=1}^{b}m_i \leq O_{p}$, $o_{p}$ should be minimized as much as possible, i.e., $o_{p}=0$. This means the optional workload of parent task must be discarded.
    \item if $\sum_{i=1}^{b}m_i > O_{p}$, $o_{p}$ should be maximized as much as possible, i.e., $o_{p}=O_{p}$. This means that the parent task should be executed precisely. A large number of child tasks and/or high values of their $m_i$ values lead to a higher chance of this scenario occurring.
\end{enumerate}

\textbf{Base Case 2: }Consider the task graph demonstrated in Fig.~\ref{fig.base_cases}b. It \textcolor{black}{consists} of a child task $c$, alongside $b$ parent tasks. The workload of \eqref{equation.bad_workload} for this simple task graph can be written as follows:
\begin{equation} \label{equation.sink_initial}
\begin{split}
    &[\sum_{i=1}^{b}(M'_i+o_i)] + \\
    &[M_{c} + m_{c} \times min\left(1,\sum_{i=1}^{b} (1-\frac{o_i}{O_i})\right)],
\end{split}
\end{equation}
in which subscripts $c$ and $i$ are used for referring to workload components of the child task and parent tasks in Fig. \ref{fig.base_cases}b, respectively.
\eqref{equation.sink_initial} can be rewritten as:
\begin{equation} \label{equation.sink}
\begin{split}
    &[\sum_{i=1}^{b}M'_i+M_{c}] + \\ 
    &[\sum_{i=1}^{b}o_i+m_{c} \times min\left(1,\sum_{i=1}^{b}(1-\frac{o_i}{O_i})\right)].
\end{split}
\end{equation}
In \eqref{equation.sink}, the first term in the summation does not depend on how many processor cycles are assigned to $o_1, o_2, ..., o_b$. 
However, the second term is a function of how many processor cycles are assigned to optional workloads of parent tasks and therefore, this term should be minimized for minimizing \eqref{equation.sink}. 
Two possible scenarios are postulated in this case:
%\textcolor{blue}{In this case, For each parent i, two possible scenarios are postulated:}
\begin{enumerate}
    \item If $\sum_{i=1}^{b}O_i \leq m_{c}$, in order to minimize \eqref{equation.sink}, all optional workloads of $b$ parent tasks should be executed completely, i.e., $\sum_{i=1}^{b}o_i=\sum_{i=1}^{b}O_i$. \textcolor{black}{The proof is beyond the scope of this paper}.
    \item If $\sum_{i=1}^{b}O_i > m_{c}$, in order to minimize \eqref{equation.sink}, all optional workloads of $b$ parent tasks should be discarded, i.e., $\sum_{i=1}^{b}o_i=0$. \textcolor{black}{The proof is beyond the scope of this paper}.
    A large number of parent tasks and/or high values of their $O_i$ values lead to a higher chance of this scenario occurring.
\end{enumerate}

\textbf{General Task Graphs:}
While base cases 1 \& 2 help determine the number of processor cycles assigned to optional workload of tasks in simple task graphs, similar conclusions cannot be drawn for complicated tasks graphs with interdependence of tasks. 
For instance, consider an example where two parent tasks share a few child tasks and the goal is to either fully discard or execute the optional workload of tasks within this task graph. 
Because a few child tasks are \textcolor{black}{potentially} shared between the two parent tasks, applying base case 1 or base case 2 without considering the interdependence of tasks may lead to conflicting decisions about execution of optional workloads. As the number of such parent tasks increases, depending on the interdependencies among them and their shared child tasks, the number of possible permutations that should be explored in terms of fully executing or discarding the optional workloads of those parent tasks can grow exponentially.  
However, presented base cases can guide us in developing a heuristic that determines the number of processor cycles assigned to optional workload of non-exit tasks. 

Note that in the proposed heuristic, it is assumed that the input task graph has only one source task (i.e. a task with in-degree of zero), but potentially many exit tasks. In task graphs where the number of source tasks is larger than one, a dummy task with zero workload is introduced and connected to all source tasks. The steps of proposed heuristic are as follows: 

%\textcolor{red}{One of the general issues I see with your writing is including super-long sentences each having multiple segments separated with commas. I suffer from the same issue in my own writings, but probably to a lesser extent. However, the good thing about this style of writing is that each sentence between commas can be easily translated to a for/if statement in pseudo-code. I just tried to rewrite those sentences to explain their main points in shorter, plain language. You can go back to the previous version and try to extract a pseudo-code for the presented heuristic and write it in a LaTeX algorithm block. This helps reviewer understand that you have thought about your algorithm clearly and were able to summarize it neatly in an algorithm block. Hopefully, this will increase the chances of getting this paper accepted. I understand that this takes some extra time, but I think it is worth it. I have personally spent more time (as a second author) on this paper compared to the time I have spent on  some of my first authored papers. But I think it is generally a good idea to make sure to publish a paper only if it shows we have put our best effort into preparing it.}

    \textbf{Step 1 (Forward Pass):} This step starts traversing tasks in the task graph $G$ from the source task and labels each task as precise (fully executing its optional workload) or imprecise (fully discarding its optional workload) based on the task's optional workload and the total maximum extension of its child tasks if the task is executed imprecisely. This step of the proposed heuristic is similar to base case 1. 
    The difference, though, is the fact that if a child task is encountered more than once due to being a shared child of multiple parent tasks and its mandatory part is extended because one of its parents is labeled as imprecise, it is not considered when writing \eqref{equation.source} for its other parent tasks. 
    After exploring all paths in the task graph, tasks with multiple parents and extended workloads are marked. 
    For these tasks, their parent tasks are evaluated again while their marked child tasks are removed from \eqref{equation.source}. 
    This may lead to an update in deciding whether the parent task should be executed precisely or imprecisely. 
    The same process is repeated until no decisions are further updated. 
    Note that each child task with multiple parents is visited only once during this update pass. %\textcolor{blue}{say about maximally extended here and/or base cases?} 

    \textbf{Step 2 (Backward Pass):} This step starts traversing tasks in the task graph $G$ in the reverse order from exit tasks back to the source task. 
    For a task with multiple parents, those which are labeled as precise are added to a list and sorted in increasing order of the number of child tasks with \textcolor{black}{intact} (not extended) mandatory workloads.
    The resulting list is called \textit{sorted\_precise\_parents}, which includes $b$ tasks. 
    Next, a subset of tasks in \textit{sorted\_precise\_parents} is chosen such that transforming those tasks to imprecise tasks and extending the mandatory workload of their child tasks leads to the highest reduction in \eqref{equation.bad_workload}. 
    However, instead of exploring all $2^{b}$ possible subsets, we only explore $b$ subsets which are: the subset containing the first task in the sorted list, the subset containing the first and second tasks in the sorted list, ..., and for the $b^{th}$ subset, the subset containing all tasks in the sorted list. 
    The rationale behind such decision is that according to base case 1, labeling a task with fewer number of intact child tasks as imprecise is more likely to eventually increase QoS. 
    Such tasks are explored more often in proposed subsets due to the sorting strategy. 

    Step 2 (Backward Pass) is inspired by base case 2 where multiple parents with shared child tasks can be labeled as imprecise. In other words, the first step of proposed heuristic looks at parent tasks independently while the second step studies their combined effect on overall QoS.

    The presented heuristic determines which tasks in a given task graph should be executed imprecisely. Therefore, we refer to this heuristic as \textit{imp\_label}. The optional workload of each non-exit task $u$ marked as imprecise is $o^{imp\_label}_u=0$ while the optional workload of a precise task is $o^{imp\_label}_u=O_u$. Furthermore, if a non-exit task $u$ has a parent which is labeled imprecise, $M'^{~imp\_label}_u=M_u+m_u$, otherwise $M'^{~imp\_label}_u=M_u$. 
    Therefore, the total workload of each non-exit task $u$ is determined by \textit{imp\_label}, is represented by $W^{imp\_label}_u$, and obtained as follows:   
    \begin{equation}
        W^{imp\_label}_u = M'^{~imp\_label}_u + o^{imp\_label}_u.
    \end{equation}
    Note that \textit{imp\_label} also determines whether the mandatory workload of an exit task $v$ is extended ($M'^{~imp\_label}_v=M_v+m_v$) or not ($M'^{~imp\_label}_v=M_v$).

\subsection{Scheduling Tasks on an MPSoC for Maximizing QoS Subject to Energy and Deadline Constraints.} \label{section.sched}
In this section, we seek to schedule the task graph obtained from \textit{imp\_label} on an MPSoC platform for maximizing QoS subject to energy and time constraints. 
For this purpose, we determine a proper processor assignment for each task alongside the ordering of tasks on each processor in order to minimize the finish time while operating at the maximum clock frequency (we temporarily ignore energy budget constraint).
This is achieved by deploying a minimal-delay list scheduling algorithm, which is a variant of Heterogeneous Earliest Finish Time (HEFT) \cite{heft}. 
HEFT assigns a rank to each task in the task graph based on the length of the critical path from that task to exit tasks. 
While HEFT is designed for heterogeneous platforms, it can be applied to a homogeneous platform as well. 
For HEFT, we provide workloads obtained from \textit{imp\_label} for non-exit tasks and for exit tasks, their (extended) mandatory workloads obtained from \textit{imp\_label} plus \textcolor{black}{their total} optional workloads. 
Next, we pick tasks in decreasing order of their ranks and schedule each selected task on its ``best'' processor, which is the processor that minimizes the finish time of the task under the maximum available frequency.
%

%As a result, HEFT enables obtaining a processor assignment for each task alongside the ordering of tasks on each processor when operating at maximum clock frequency.
%\textcolor{blue}{However, the number of processor cycles assigned to optional workload of exit tasks, the operating clock frequency, and the execution start time of each task need to be found and optimized using a different algorithm.}
Note that HEFT is only used to just obtain a processor assignment for each task alongside the ordering of tasks on each processor. The obtained start times for tasks from HEFT just show relative ordering of tasks on each processor. Furthermore, we used the maximum frequency in HEFT and included the total optional workloads of all exit tasks since we were temporarily ignoring the energy budget constraint. Therefore, in the next step, the actual number of processor cycles assigned to optional workload of exit tasks, the actual distribution of workload of each task among $m$ available frequencies of the processors, and the actual execution start time of each task should be obtained.

%For this purpose, we demonstrate that maximizing QoS for a task graph obtained from \textit{imp\_label} subject to energy and time constraints will be reduced to a linear programming (LP) formulation if HEFT is applied \textcolor{blue}{a priori} to find processor assignment and task ordering. In this formulation, $u$ and $v$ represent a task in the task graph.
For this purpose, we demonstrate that maximizing QoS for a task graph obtained from \textit{imp\_label} subject to energy and time constraints, and processor assignment and task ordering obtained from HEFT, will be reduced to a linear programming (LP) formulation. \textcolor{black}{In the following formulation, $u$ and $v$ are used to refer to any of the tasks in the task graph}.
%\textcolor{blue}{should we say whether exit or non-exit task?} 
%

Duration of task $u$, $u=1,2, ..., n$, is formulated as follows:
\begin{equation} \label{equation.duration}
    D_u = \sum_{i=1}^{m} \frac{N_{u,i}}{f_i}, \quad N_{u,i} \geq 0
\end{equation}
where $N_{u,i}$ indicates the number of processor cycles of task $u$ processed at clock frequency $f_i$ ($i = 1, 2, ..., m$). If task $u$ is a non-exit task,  the following constraint is introduced:
\begin{equation} \label{equation.non-exit_workload_cons}
    \sum_{i=1}^{m} N_{u,i}=W^{imp\_label}_u.
\end{equation}
On the other hand, if task $u$ is an exit task, we have:
\begin{equation} \label{equation.exit_workload_cons}
    M'^{~imp\_label}_u \leq \sum_{i=1}^{m} N_{u,i} \leq M'^{~imp\_label}_u + O_u.
\end{equation}

According to \eqref{equation.power} and \eqref{equation.duration}, energy consumption during the execution of task $u$ can be formulated as follows:
\begin{equation} \label{equation.task_energy}
    \epsilon_{task}(u) = \sum_{i=1}^{m}(N_{u,i}.(\alpha f_i^{\beta-1}+\gamma+\frac{\delta}{f_i})).
\end{equation}
%\textcolor{red}{This is what I was talking about: using multiple letters for a single variable would potentially cause confusion. Don't assume that reviewers will read your paper with 100\% concentration. They will probably read it during their coffee berak.}
%\begin{equation}
%    N_{u,i} \geq 0.
%\end{equation}
To ensure the total energy consumption of tasks is less than or equal to the given energy bound, represented by $\epsilon_{max}$, we have:
\begin{equation}
    \sum_{u=1}^{n} \epsilon_{task}(u) \leq \epsilon_{max}.
\end{equation}
%For calculating total energy consumption, we assume processors can be put in deep sleep mode without time and energy overheads when not executing a task (the model can be easily extended to incorporate this cost). \textcolor{blue}{should we still say this?}

To ensure time and precedence constraints, by representing start time of each task $u$ with $S_u$, we should have:
\begin{equation} \label{equation.Td}
    S_u + D_u \leq T_d, \quad u = 1, 2, ..., n,~S_u \geq 0,
\end{equation}

\begin{equation} \label{equation.precedence}
    S_u + D_u + \overline{C_{u,v}}\leq S_v, \quad \forall e(u,v) \in E.
\end{equation}
In \eqref{equation.precedence}, $\overline{C_{u,v}}$ represent the average communication cost associated with $e_{u,v}$ for sending output of task $u$ to input of task $v$.
\iffalse
To ensure each task finishes its execution before $T_d$, for each of $n$ tasks we have:
\begin{equation} \label{equation.Td}
    S_u + Dur_u \leq T_d, \quad S_u \geq 0,
\end{equation}
%\begin{equation}
%    S_u \geq 0,
%\end{equation}
where $S_u$ represents start time of the execution of task $u$. Furthermore, for precedence constraint, we have:
\begin{equation} \label{equation.precedence}
    S_u + Dur_u \leq S_v, \quad \forall e(u,v) \in E.
\end{equation}
Here, we do not consider any inter-task communication cost associated with $e(u,v)$ for sending output data of task $u$ to input data of task $v$ (The model can be easily extended to incorporate this cost).
%\textcolor{red}{?}.
\fi

Finally, we need to ensure tasks assigned to the same processor do not overlap:
\begin{equation} \label{equation.nonpreemptive}
\begin{split}
    S_u + D_{u} \leq S_v, ~ &\textrm{For tasks $u$ and $v$ which are} \\
    &\textrm{assigned to the same processor} \\
    &\textrm{and task $v$ is the immediate task} \\
    & \textrm{after task $u$ based on HEFT}
\end{split}
\end{equation}

Maximizing the objective function of \eqref{equation.QoS}, with the constraints introduced in \eqref{equation.duration} to \eqref{equation.nonpreemptive}, forms an LP over positive real variables of $S_u$, $N_{u,i}$, and optional workload of exit tasks ($o_u$ for $u$ $\in$ exit tasks). %\textcolor{red}{I only have used "set" notation for exit tasks in equation (8) and not other places to avoid confusion (such as here or equation (10)). Is that okay?}

\subsection{\textcolor{black}{MILP formulation}} \label{MILP}
In order to evaluate the performance of our 2-step proposed method in Sections \ref{section.imp_label} and \ref{section.sched} compared to the optimal solution, we present a comprehensive mixed-integer linear programming (MILP) formulation of the problem statement in Section \ref{subsection.problem_statement}. By solving the MILP, we obtain the optimal values for the number of processor cycles assigned to the optional workload of each task, processor assignment for each task alongside the ordering of tasks on each processor, task execution start time, and distribution of the total number of processor cycles associated with the execution of each task among $m$ available frequencies. For this purpose, the following variables are defined:

Denoting the number of processors with $K$, for the processor assignment of task $u$ to processor $k$, $k=1,2, ..., K$, we use the decision variable $\Pi_{k,u}$, defined as follows:
\begin{equation}
        \Pi_{k,u} = \left\{ \begin{array}{cc} 
                1 & \hspace{5mm} \textrm{if task u is assigned to processor k} \\
                0 & \hspace{5mm} \textrm{otherwise} \\
                \end{array} \right..
\end{equation}
Consequently, we have the following constraint for $\Pi_{k,u}$:
\begin{equation} \label{cons_processor}
        \sum_{k=1}^{K} \Pi_{k,u}=1, \quad \textrm{for } u = 1, 2, ..., n.
\end{equation}

In order to prevent the overlap of execution of tasks assigned to the same processor with each other, we use the decision variable $Y_{k,u,v}$ indicating ordering of the tasks. For $k = 1, 2, ..., K$; $u = 1, 2, ..., n$; $v = 1, 2, ..., n, v \neq u;$ we define:
\begin{equation} \label{equation.O}
    Y_{k,u,v} = \left\{ \begin{array}{ll} 
                1 & \hspace{1mm} \textrm{if task u is scheduled immediately} \\
                 & \hspace{1mm} \textrm{ before task v on processor k} \\
                 & \hspace{1mm}  \\
                0 & \hspace{1mm} \textrm{otherwise} \\
                \end{array} \right..
\end{equation}
In addition, if task $v$ is the first task assigned to processor $k$, $Y_{k,0,v}$ is defined to be 1 (and is 0 otherwise). On the other hand, if task $u$ is the last task assigned to processor $k$, $Y_{k,u,n+1}$ is defined to be 1 (and is 0 otherwise).
%Furthermore, we define $O_{k,0,n+1}$ which is equal to 1 when there is no task assigned to processor $k$ (and is 0 otherwise).
Furthermore, if there is no task assigned to processor $k$, $Y_{k,0,n+1}$ is defined to be 1 (and is 0 otherwise). 
Accordingly, using \eqref{equation.O} and the definitions provided for $Y_{k,0,v}$, $Y_{k,u,n+1}$ and $Y_{k,0,n+1}$, we have the following constraints for $k = 1, 2, ..., K$:
%\begin{equation} \label{equation.next}
%    \sum_{\substack{v=1 \\ v\neq u}}^{n+1} O_{k,u,v} = \Pi_{k,u}, \quad \textrm{for } \substack{u = 0, 1, %..., n\\ k = 1, 2, ..., K} 
%\end{equation}
\begin{equation} \label{equation.next}
     \sum_{\substack{v=1 \\ v\neq u}}^{n+1} Y_{k,u,v} = \Pi_{k,u}, \quad \textrm{for } u = 0, 1, ...,n 
\end{equation}
%\begin{equation} \label{equation.prev}
%    \sum_{\substack{u=0 \\ u\neq v}}^{n} O_{k,u,v} = \Pi_{k,v}. \quad \textrm{for } \substack{v = 1, 2, %..., n+1\\ k = 1, 2, ..., K} 
%\end{equation}
\begin{equation} \label{equation.prev}
    \sum_{\substack{u=0 \\ u\neq v}}^{n} Y_{k,u,v} = \Pi_{k,v}, \quad \textrm{for } v = 1, 2, ..., n+1.
\end{equation}
According to (\ref{equation.next}), if task $u$ is assigned to processor $k$ ($\Pi_{k,u}=1$), either there is one and only one task scheduled immediately after task $u$ on processor $k$ or task u is the last task assigned to processor $k$. Similarly, according to (\ref{equation.prev}), if task $v$ is assigned to processor $k$ ($\Pi_{k,v}=1$), either there is one and only one task scheduled immediately before task $v$ on processor $k$ or task $v$ is the first task assigned to processor $k$. In both (\ref{equation.next}) and (\ref{equation.prev}), $\Pi_{k,0}$ and $\Pi_{k,n+1}$ are defined as 1 for all $k = 1, 2, ..., K$. Using $Y_{k,u,v}$, we rewrite the constraint in \eqref{equation.nonpreemptive} as the following:
\begin{equation} \label{equation.nonpreemptive_milp}
\begin{split}
    S_u + D_u - (1 - Y_{k,u,v}) \times T_d \leq S_v, \\
    \textrm{for } u = 1, 2, ..., n, \\
    \textrm{for } v = 1, 2, ..., n, v \neq u, \\
    \textrm{for } k = 1, 2, ..., K.
\end{split}
\end{equation}

Finally, instead of using \textit{imp\_label} algorithm to determine the workload of non-exit and exit tasks in \eqref{equation.non-exit_workload_cons} and \eqref{equation.exit_workload_cons}, the following constraint is used for all the tasks:
\begin{equation} \label{equation.MILP_task_workload_cons}
    M_u + m_u \times E^i_u \leq \sum_{i=1}^{m} N_{u,i} \leq M_u + m_u \times E^i_u + O_u,
\end{equation}
where $E^i_u$ is obtained by \eqref{equation.input_error}. In order to present the minimum formulation existing in \eqref{equation.input_error} as a linear constraint, we rewrite \eqref{equation.input_error} using an auxiliary decision variable, represented by $X_u$, as the following:
\begin{equation} \label{eq_input_error_X}
    E^i_u=X_u.\left(1\right)+(1-X_u).(\sum_{j \in par(u)}{E^o_j}),
\end{equation}
in which $X_u$ is a decision variable which is 1 when $\sum_{j \in par(u)}{E^o_j} > 1$ and is 0 otherwise. According to \cite{esmaili}, the corresponding constraint for $X_u$ can be written as follows:
\begin{equation} \label{cons_x}
    \frac{\sum_{j \in par(u)}{E^o_j} - 1}{n} \leq X_u \leq \sum_{j \in par(u)}{E^o_j}, \quad X_u \in \{0,1\},
\end{equation}
in which $n$ serves as an upper bound for $\sum_{j \in par(u)}{E^o_j}$. Furthermore, we use the lemma presented in \cite{esmaili} for linearization of multiplication of a Boolean decision variable and a bounded real-valued variable for the second term of \eqref{eq_input_error_X}. 

Consequently, maximizing the objective function of \eqref{equation.QoS} with the constraints introduced in \eqref{equation.duration}, \eqref{equation.task_energy} to \eqref{equation.precedence}, \eqref{cons_processor}, \eqref{equation.next} to \eqref{cons_x}, and the lemma mentioned in \cite{esmaili} for linearization of the second term of \eqref{eq_input_error_X}, forms an MILP yielding the optimal values for the desired variables mentioned in the beginning of this section.

\subsection{Complexity Analysis}
 The time complexity of the proposed labeling heuristic described in Section \ref{section.imp_label} is $\mathcal{O}(|E| + |V|)$ where $|E|$ denotes the number of edges in the task graph while $|V|$ represents the number of vertices. Furthermore, the time complexity of HEFT, which is used for obtaining the processor assignment of tasks in the labeled graph and ordering of them on each processor for an MPSoC platform, is $\mathcal{O}(K \times |E|)$ where $K$ denotes the number of processors. %\textcolor{blue}{Consequently, the overall complexity of the proposed heuristic for scheduling a task graph with potentially imprecise computations subject to an energy bound and a hard deadline on an MPSoC platform is $\mathcal{O}(K \times |E| + |V|)$.}

\section{Results} \label{section.results}
\subsection{\textcolor{black}{Simulation Setup}} \label{simu_setup}
For solving the formulated MILP in Section \ref{MILP} and the LP part of the porposed method in Section \ref{section.sched}, we use IBM ILOG CPLEX Optimization Studio\textcolor{black}{\cite{cplex}}. The platform on which simulations are performed is a computer with a 3.2 GHz Intel Core i7-8700 Processor and 16 GB RAM. For obtaining energy model parameters, we employ \cite{chen2014energy} which uses a classical energy model of a 70nm technology processor that supports 5 discrete frequencies. The frequency-independent component of processor power consumption, which is represented by $\delta$ in (\ref{equation.power}), is obtained as $276\,mW$. Each processor can operate independently of other processors at either $f_1=1.01\,GHz$, $f_2=1.26\,GHz$, $f_3=1.53\,GHz$, $f_4=1.81\,GHz$, $f_5=2.1\,GHz$. For these frequencies, frequency-dependent component of processor power consumption, which is represented by $\alpha f^\beta+\gamma f$ in (\ref{equation.power}), is $430.9\,mW$, $556.8\,mW$, $710.7\,mW$, $896.5\,mW$, and $1118.2\,mW$, respectively. Using curve fitting, we obtain $\alpha=23.8729$, $\gamma=401.6654$, and $\beta=3.2941$ in (\ref{equation.power}). We consider an architecture with 4 processors. Simulations are performed for \textcolor{black}{10} task graphs randomly generated using TGFF\textcolor{black}{\cite{tgff}}, which is a randomized task graph generator widely used in the literature to evaluate the performance of scheduling algorithms. These task graphs are named as TGFF0 to TGFF9. The number of tasks in studied random task graphs ranges from \textcolor{black}{23 (in TGFF0) to 57 (in TGFF9)}. The maximum in-degree and out-degree for each task in our randomly generated task graphs are set to 6. 
%For each task $u$, the workload of its base mandatory workload ($M_u$), plus the maximum workload of its optional part ($O_u$), is the amount of workload required to be assigned to the task when input is error-free in order to produce completely precise results. we refer to this workload as initial workload of the task, which is represented by $W^{initial}_u$. Therefore: $W^{initial}_u = M_u + O_u$.
For each task $u$, the amount of workload required to be assigned to produce precise results when input is error-free is referred to as the initial workload of the task, and is represented by $W^{initial}_u$. Therefore: $W^{initial}_u = M_u + O_u$. For studied task graphs, the average value for $W^{initial}_u$ of each task $u$ is set to \textcolor{black}{$2 \times 10 ^ 6$} cycles. For each task $u$, based on what portion of $W^{initial}_u$ is for its base mandatory workload ($M_u$), we consider 3 cases: 
\begin{enumerate}
\item \textcolor{black}{$man\_low$}: $M_u \sim U(0.2,~0.4) \times W^{initial}_u$ (low portion of $W^{initial}_u$ is \textcolor{black}{for the base mandatory}).
\item $man\_med$: $M_u \sim U(0.4,~0.6) \times W^{initial}_u$ (medium portion of $W^{initial}_u$ is for the base mandatory).
\item $man\_high$: $M_u \sim U(0.6,~0.8) \times W^{initial}_u$ (high portion of $W^{initial}_u$ is for the base mandatory).
\end{enumerate}
\textcolor{black}{In each of 3 cases, similar to\cite{Greece2010}, $m_u$ is set as $m_u \sim U (0,~2 \times M_u)$ for each task $u$}. For having a fair comparison among these 3 cases, each task graph uses the same random seed for all the above uniform distributions, where this random seed is different in each task graph. In all 3 cases, $P^T_u$ for all the tasks are uniformly chosen from [0, 1]. Average communication costs associated with edges of task graphs are chosen uniformly from 0.4 ms to 0.6 ms. $T_d$ of each task graph is set to twice the length of the longest path from its source task to an exit task (including communication costs), when executing the total workload along the path, including all optional workloads, with the maximum frequency. 

\begin{figure*}[b]
\centering
%\captionsetup{justification=centering}
\includegraphics[width=0.99\textwidth]{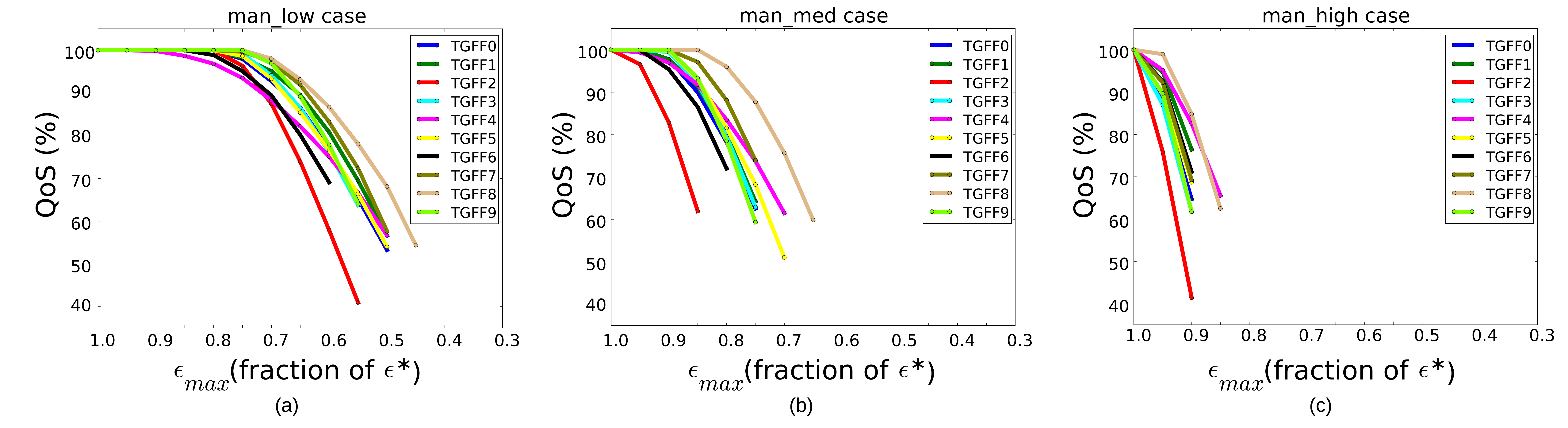}
\caption{\textcolor{black}{ \footnotesize QoS versus $\epsilon_{max}$ obtained via the proposed method for different cases of mandatory workload portion: (a) man\_low (b) man\_med (c) man\_high.}} \label{fig.man_cases}
\end{figure*}

\subsection{Evaluating the Effect of Energy Budget on the obtained QoS}
In this section, for each of the studied task graphs, we evaluate the effect of the $\epsilon_{max}$ value on the obtained QoS, defined in \eqref{equation.QoS}, using the proposed method in Sections \ref{section.imp_label} and \ref{section.sched}. In order to obtain a proper value for $\epsilon_{max}$, first, we derive the minimum energy required for scheduling the task graph in one $T_d$ without the possibility of imprecise computations. We refer to this energy value as $\epsilon^*$.
%\textcolor{purple}{For obtaining $\epsilon^*$, we solve the MILP which minimizes the objective function of $\sum_{u=1}^{n}\epsilon_{task}(u)$, with constraints in \eqref{equation.duration}, \eqref{equation.task_energy}, \eqref{equation.Td}, \eqref{equation.precedence}, \eqref{cons_processor}, \eqref{equation.next} to \eqref{equation.nonpreemptive_milp}, and the constraint imposing that the workload of each task $u$, whether non-exit task or exit task, should be executed precisely: $\sum_{i=1}^{m} N_{u,i}=W^{initial}_u$. By solving this MILP, $\epsilon^*$ of each task graph will be obtained.}  
For obtaining $\epsilon^*$, HEFT is again used to obtain the processor assignment for each task and the ordering of tasks on each processor. Then, we solve the LP which minimizes the objective function of $\sum_{u=1}^{n}\epsilon_{task}(u)$, with the constraints described in \eqref{equation.duration}, \eqref{equation.task_energy}, \eqref{equation.Td} to \eqref{equation.nonpreemptive}, and the constraint imposing that the workload of each task $u$, whether non-exit task or exit task, should be executed precisely: $\sum_{i=1}^{m} N_{u,i}=W^{initial}_u$. By solving this LP, $\epsilon^*$ of each task graph will be obtained. Optionally, one can use an MILP formulation to obtain  $\epsilon^*$.

%\begin{figure}[!t]
%\centering
%%\captionsetup{justification=centering}
%\includegraphics[width=0.4\textwidth]{./figures/base_man_1.pdf}
%\caption{\textcolor{black}{ \footnotesize The QoS versus $\epsilon_{max}$ obtained via the baseline approach (without the \textit{imp\_label} algorithm) for the man\_low case}} \label{fig.base_man_1}
%\end{figure}

For the case of imprecise computations, for each task graph, if its $\epsilon^*$ is used as the value for $\epsilon_{max}$, QoS is obtained as its maximum value (QoS = 100\%, if QoS in \eqref{equation.QoS} is shown with percentage). Therefore, for each task graph, we reduce $\epsilon_{max}$ gradually, starting from its $\epsilon^*$ with the resolution of $0.05 \times \epsilon^*$, and observe QoS obtained using our proposed method for each value of $\epsilon_{max}$, as presented in Fig. \ref{fig.man_cases}. In Fig. \ref{fig.man_cases}, for each task graph, existence of a QoS $\geq$ 0 for a ratio of its $\epsilon^*$ as $\epsilon_{max}$, shows that our proposed method can generate a feasible schedule for that task graph and $\epsilon_{max}$, which produces that value of QoS. A feasible schedule manes at least (extended) mandatory workloads of all tasks are completed before $T_d$, and the total energy consumption is below the $\epsilon_{max}$.

According to Fig. \ref{fig.man_cases}, by reducing $\epsilon_{max}$, we observe the sharpest drop in the obtained QoS by our proposed method in \textcolor{black}{the} $man\_high$ case, while the slowest drop in QoS is observed in \textcolor{black}{the} $man\_low$ case. This reflects the fact that when lower portion of initial task workloads are mandatory, feasible results can be achieved with lower values of $\epsilon_{max}$, compared to the case that higher portion of initial task workloads are mandatory. For instance, in \textcolor{black}{the} $man\_low$ case, our proposed method can generate a feasible schedule for TGFF8 even with using 45\% of its $\epsilon^*$ as the value for $\epsilon_{max}$, while in \textcolor{black}{the} $man\_high$ case, it can only generate a feasible schedule for TGFF8 when $\epsilon_{max}$ is reduced to at most 85\% of its $\epsilon^*$.

\subsection{Evaluating the Performance of the Proposed Method versus MILP} \label{section.method_performance}
\begin{figure} [b]
    \centering
    \begin{subfigure}[t]{0.45\textwidth}
        \centering
        \includegraphics[width = 1\textwidth]{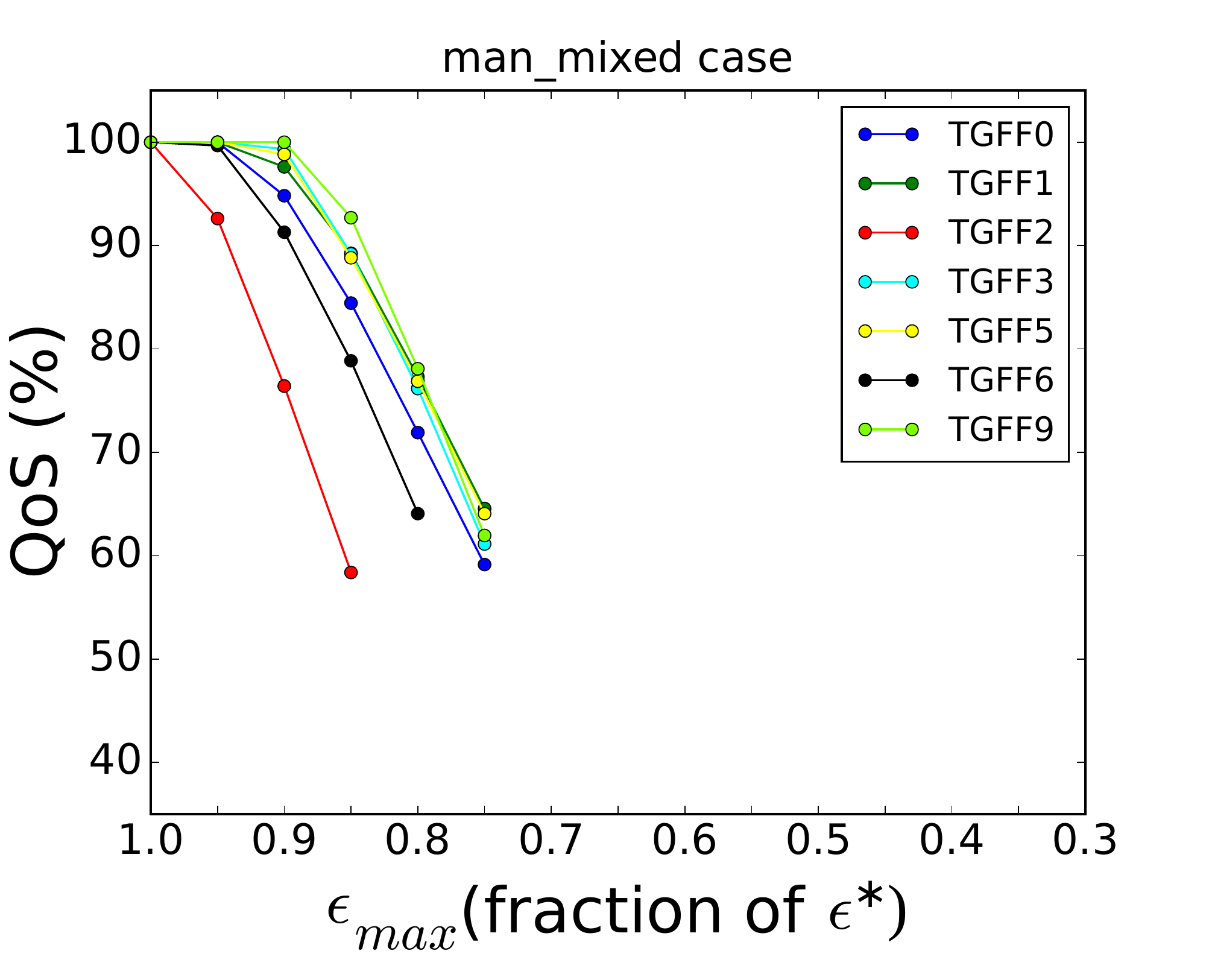}
        \caption{}
        \label{fig.bland_PTTT}
    \end{subfigure}
    \begin{subfigure}[t]{0.45\textwidth}
        \centering
        \includegraphics [width = 1\textwidth]{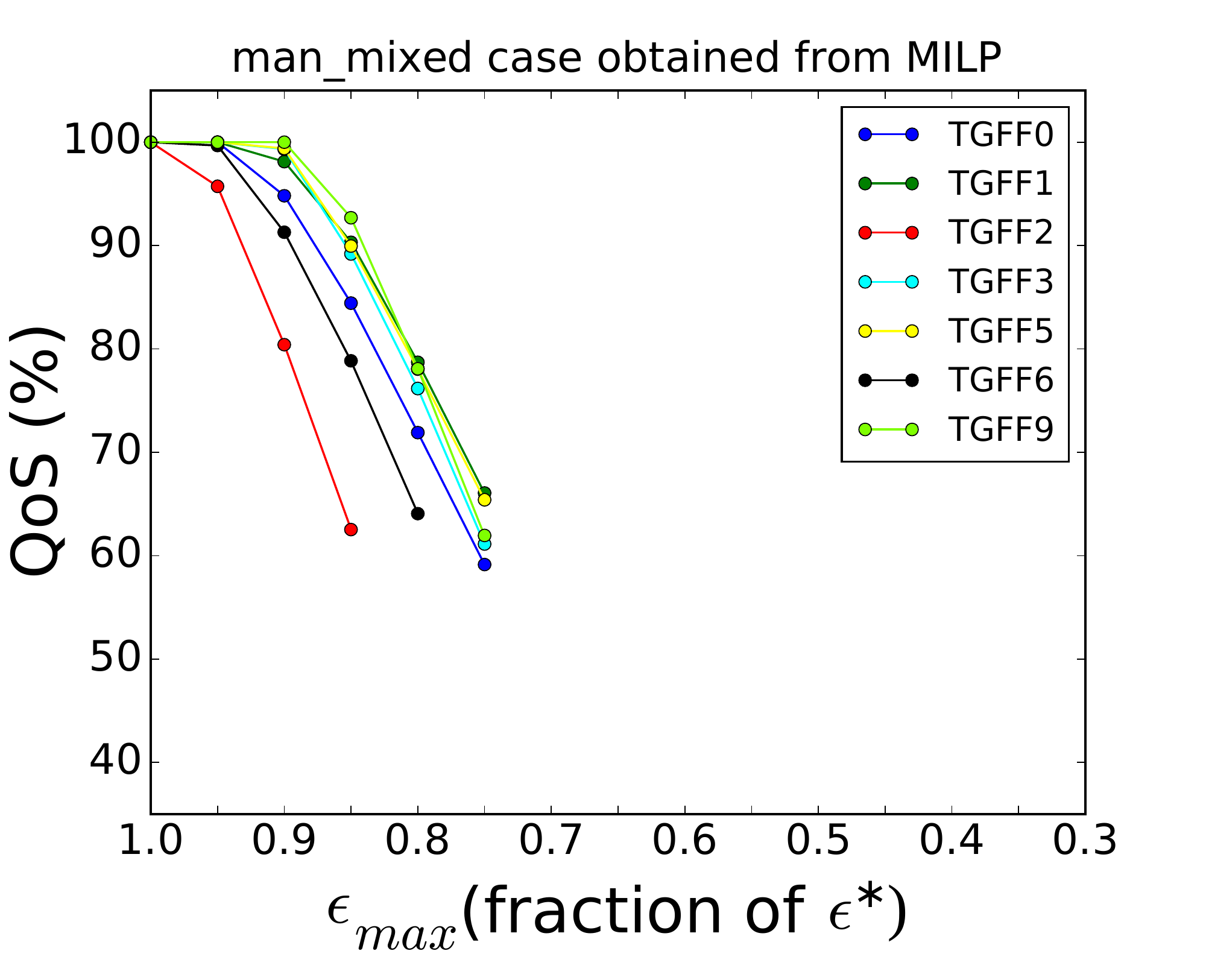}
        \caption{}
        \label{fig.bland_PATT}
    \end{subfigure}
    \caption{\footnotesize{QoS versus $\epsilon_{max}$ obtained via (a) the proposed method and (b) MILP for the $man\_mixed$ case. As presented, QoS values obtained using the proposed method are close (and in some cases completely equal) to the optimal reference QoS values found by MILP.}}\label{fig.method_milp}
\end{figure}

In this section, we compare the the performance of the proposed method in Sections \ref{section.imp_label} and \ref{section.sched} with the MILP formulation presented in Section \ref{MILP}, in terms of their obtained QoS in different values of $\epsilon_{max}$. We consider our comparison in a case where $M_u$ of tasks in a task graph can be chosen uniformly from 20\% to 80\% of $W_u^{initial}$ (a mix of 3 aforementioned cases in Section \ref{simu_setup}; we refer to this case as the $man\_mixed$ case). For each task graph and $\epsilon_{max}$ value, we impose a time limit of 60 minutes for MILP to find the optimal scheduling solutions. For evaluating the performance of our proposed method, we only consider those task graphs for which MILP found the optimal solutions for each value of $\epsilon_{max}$ within the time limit (This comparison is actually in the favor of MILP. We elaborate more on this later). Using this setup, MILP could find solutions for 7 of 10 studied task graphs within the time limit (TGFF0 to TGFF3, TGFF5, TGFF6 , and TGFF9). For these task graphs, the obtained QoS values for different $\epsilon_{max}$ values by the proposed method and MILP are shown in \textcolor{black}{Fig. \ref{fig.method_milp}}. According to \textcolor{black}{Fig. \ref{fig.method_milp}}, QoS values found by our proposed method are completely equal to those found by MILP for 4 task graphs (TGFF0, TGFF3, TGFF6, and TGFF9). Furthermore, for other task graphs, the average QoS difference found by the proposed method versus MILP for different $\epsilon_{max}$ values is 1.63\% (up to 6.64\%). Consequently, the proposed method yeilds close QoS values compared to the optimal MILP formulation.

%\textcolor{blue}{say a paragraph about time advantage of our method versus MILP and the fact that MILP fails for some cases within the time limit and the fact that the number of nodes goes higher complexity of MILP goes higher exponentially and our method would be better then}
Comparing the runtime of the proposed method and MILP, we see a clear advantage for the proposed method. On the platform we performed simulations, the average runtime of the proposed method for each task graph and $\epsilon_{max}$ value was 99.38\% lower compared to MILP. This is without considering the cases that MILP did not find the optimal solutions within the time limit. for many real-world applications, as the task graphs can have higher number of nodes and more complex interdependencies compared to studied task graphs, the runtime of using MILP for those task graphs can grow exponentially. Therefore, employing the proposed method, as it provided close estimations to MILP, can be an efficient alternative.     

\begin{figure} [b]
    \centering
    \begin{subfigure}[t]{0.45\textwidth}
        \centering
        \includegraphics[width = 1\textwidth]{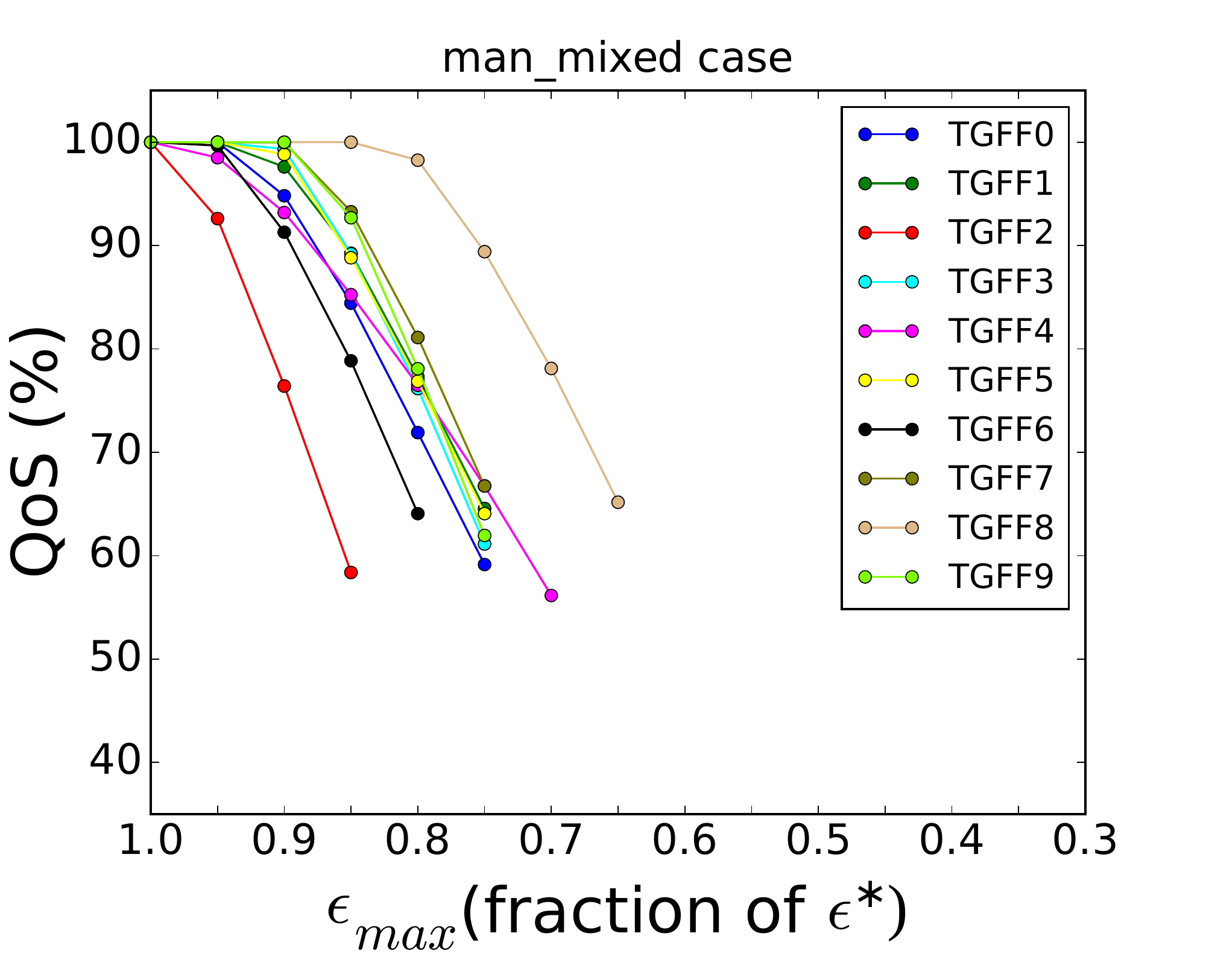}
        \caption{}
        \label{fig.bland_PTT}
    \end{subfigure}
    \begin{subfigure}[t]{0.45\textwidth}
        \centering
        \includegraphics [width = 1\textwidth]{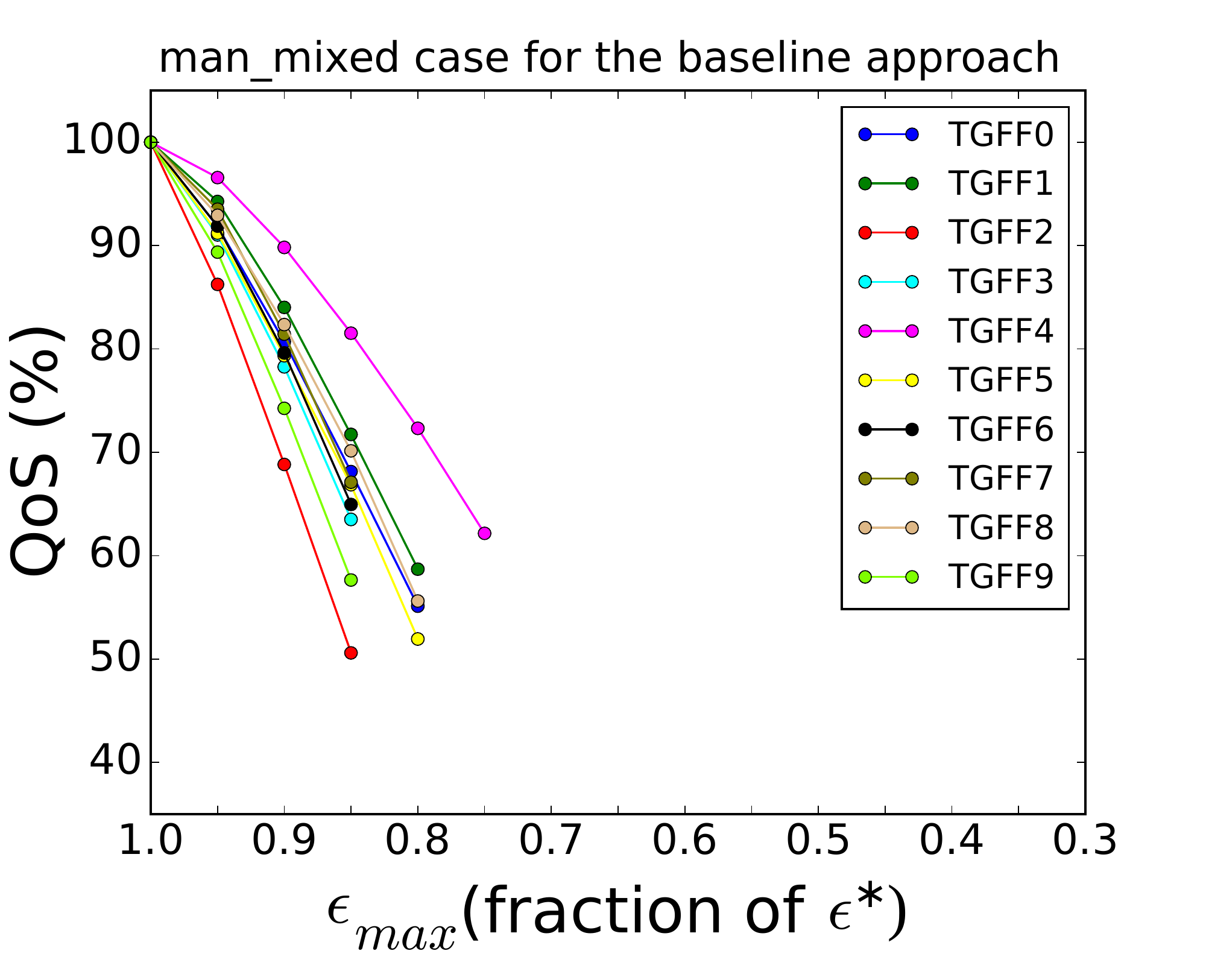}
        \caption{}
        \label{fig.bland_PAT}
    \end{subfigure}
    \caption{\footnotesize{QoS versus $\epsilon_{max}$ obtained via (a) the proposed method and (b) the baseline approach (without the \textit{imp\_label} algorithm) for the $man\_mixed$ case. As presented, QoS values obtained using the proposed method are considerably higher compared to the baseline approach.}}\label{fig.method_base}
\end{figure}

%\textcolor{blue}{The average QoS obtained from the proposed method is only 2.53\% (and up to 6.22\%) lower than the average optimal QoS obtained from the MILP formulation for different values of $\epsilon_{max}$ over all the task graphs. Consequently, we observe the proposed method yeilds close QoS values compared to the optimal MILP formulation. The obtained results are promising considering the fact that on the platform we performed simulations, solving the MILP for each task graph and $\epsilon_{max}$ value provided the optimal results on average in 96 minutes, while the proposed method provided the results in less than 2 seconds for all the cases.} 
%For illustration, we present the comparison between the proposed method and MILP for the task graph for which the difference between the proposed method and MILP was at its maximum among all task graphs in Fig. \ref{} (in our case it was TGFF3). 
%Consequently, we observe the proposed method yeilds close QoS values compared to the optimal MILP formulation, while on the platform we performed simulations, solving the MILP for each task graph and $\epsilon_{max}$ value provided the optimal results on average in 96 minutes, while the proposed method provided the results in less than 2 seconds for all the  cases. 

\subsection{Evaluating the Effect of imp\_label algorithm}
In order to evaluate the effect of \textit{imp\_label} algorithm \textcolor{black}{presented} in Section \ref{section.imp_label}, which traverses the graph and labels some tasks as the ones that should be executed imprecisely before feeding that graph to the scheduling method presented in Section \ref{section.sched}, we compare the results obtained from our proposed method with a baseline approach in which we feed the task graph with their initial workloads ($W^{initial}$) for non-exit tasks to the scheduling method presented in Section \ref{section.sched}, and assign as much as processor cycles possible to exit tasks in order to maximize QoS. Therefore, In the baseline approach, we solve the same LP as the one \textcolor{black}{formulated} in Section \ref{section.sched}, however, the constraint in \eqref{equation.non-exit_workload_cons} for non-exit task $u$ will be transformed to the following constraint:
\begin{equation} \label{equation.non-exit_workload_cons_BASE}
    \sum_{i=1}^{m} N_{u,i}=W^{initial}_u,
\end{equation}
and the constraint in \eqref{equation.exit_workload_cons} for exit task $u$ will be transformed to the following constraint:
\begin{equation} \label{equation.exit_workload_cons_BASE}
    M_u \leq \sum_{i=1}^{m} N_{u,i} \leq M_u + O_u.
\end{equation}

Fig. \ref{fig.method_base} presents QoS values obtained via the proposed method and the baseline approach for the studied task graphs for different values of $\epsilon_{max}$. The base mandatory portion of initial workload of tasks is set based on the $man\_mixed$ case, similar to Section \ref{section.method_performance}. \textcolor{black}{According to Fig. \ref{fig.method_base},} using the baseline approach, QoS for all task graphs immediately drops from 100\% as soon as $\epsilon_{max}$ is reduced from $\epsilon^*$. However, in the corresponding $man\_mixed$ case of our proposed method, as shown in Fig. \ref{fig.method_base}, QoS can be maintained as 100\% even for values lower than $\epsilon^*$ for our studied task graphs. For instance, as presented in Fig. \ref{fig.method_base}, our proposed method can generate QoS of 100\% for \textcolor{black}{TGFF8 with 85\%} of its $\epsilon^*$. Furthermore, for each task graph, the minimum $\epsilon_{max}$ with which our proposed method can generate a feasible schedule for that task graph is lower in comparison to the baseline approach. For those $\epsilon_{max}$ values that both the proposed method and the baseline approach can provide a feasible schedule for, QoS values obtained via our proposed method are on average 12.82\% (up to 43.40\%) higher than QoS values obtained via the baseline approach.

\section{Conclusion} \label{section.conclusion}
In this paper, we \textcolor{black}{presented} a method for time and energy constrained scheduling of task graphs on MPSoC platforms, with the possibility of imprecise computation of each task of the task graph. We took into the account the effect of the extension in the workload of each task when the input to that task is not precise. For this purpose, we \textcolor{black}{presented} an algorithm which by traversing the task graph, determines \textcolor{black}{the} optional workload of each non-exit task should be executed or discarded, and then scheduled the labeled graph on a MPSoC platform. For evaluating the efficacy of the proposed method, we also presented a MILP formulation of the problem which provided us the optimal reference scheduling solutions. Our results shows the effectiveness of our proposed method in terms of obtaining promising QoS values even with low energy budgets.

%\section*{Acknowledgment}

%
% The next two lines define the bibliography style to be used, and the bibliography file.
%\bibliographystyle{ACM-Reference-Format}
%\bibliography{sample-base}
\bibliographystyle{unsrt}
%\bibliography{bibliography}

\end{document}